\documentclass{PoS}

\title{The Chrial Magnetic Wave and \\ Strong Field Effects in Heavy Ion Collisions}

\ShortTitle{The Chrial Magnetic Wave}

\author{\speaker{Jinfeng Liao}%
        \thanks{The author acknowledges partial support from RIKEN BNL Research Center. He thanks his collaborators on various results reported here, including Y. Burnier, D. Kharzeev, H. Yee, X. Huang, X. Zhang, and J. Bloczynski. He is also grateful to V. Koch, N. Xu, A. Tang, G. Wang, and H. Ke for useful discussions. }\\
       Physics Department and Center for Exploration of Energy and Matter,
Indiana University, 2401 N Milo B. Sampson Lane, Bloomington, IN 47408, USA.\\
 RIKEN BNL Research Center, Bldg. 510A, Brookhaven National Laboratory, Upton, NY 11973, USA. \\
       E-mail: \email{Jinfeng Liao <liaoji@indiana.edu>}}

\abstract{A number of recent progresses in the study of strong field effects in heavy ion collisions are discussed here: 1) the Chiral Magnetic Wave (CMW) and its experimental manifestation via splitting of positive/negative pions' elliptic flow; 2) the event-by-event azimuthal fluctuations of strong EM fields and its correlations with matter geometry; 3) a new mechanism for generating axial current in external electric field, the Chiral Electric Separation Effect (CESE). }

\FullConference{8th International Workshop on Critical Point and Onset of Deconfinement\\
		 March 11-15\\
		 Napa, California, USA}

\begin{document}

\section{Introduction}

Recently there are strong interests in understanding the responses of a relativistic plasma with chiral fermions to externally applied Maxwell electromagnetic (EM) fields. The hot deconfined QCD matter known as the quark-gluon plasma (QGP), created in relativistic heavy ion collisions, is a good (approximate) example of such a plasma. Furthermore during the early stage of heavy ion collisions there are very strong EM fields originated from the fast moving protons inside nucleus. This therefore provides an environment for studying strong field effects in QCD plasma.  

In particular, it has been found that the QCD axial anomaly could induce the following two phenomena in the QGP with the presence of an external magnetic field:
the Chiral Magnetic Effect (CME) and the Chiral Separation Effect (CSE) ~\cite{Kharzeev:2004ey,son:2004tq,Metlitski:2005pr,Zhitnitsky:2012im}.
The CME is the generation of vector current and thus the electric charge separation along the axis of the applied magnetic field in the presence of nonzero axial charge density arising from fluctuating topological charge \cite{Kharzeev:2004ey}. With an imbalance
between the densities of left- and right-handed quarks, parameterized by an axial
chemical potential $\mu_A$, an external magnetic field induces the vector current $j^i_V = \langle\bar{\psi} \gamma^i \psi\rangle$:
\begin{eqnarray}
\vec j_V=\sigma_5 \mu_A \vec B; \label{cme}
\end{eqnarray}
with chiral conductivity $ \sigma_5\equiv {N_ce \over 2\pi^2}$. Such an imbalance of chirality (thus nonzero $\mu_A$) may arise in QGP via instanton and/or sphaleron transitions.  The CSE on the other hand predicts the generation of an axial current, $j_A^i=\langle\bar{\psi} \gamma^i\gamma_5 \psi\rangle$, and thus separation of axial charges along the  external $\vec B$ field at nonzero vector charge density (parameterized by its chemical potential $\mu_V$) \cite{son:2004tq,Metlitski:2005pr}:
\begin{eqnarray}
\vec j_A=\sigma_5  \mu_V \vec B . \label{cse}
\end{eqnarray}
Since there are extremely strong transient $\vec E$ and $\vec B$ fields \cite{Bzdak:2011yy,Deng:2012pc,Bloczynski:2012en,Tuchin:2010vs,Voronyuk:2011jd,McLerran:2013hla,Tuchin:2013apa} across the QCD plasma in heavy ion collisions, a lot of efforts have been made to search for such strong field effects. For example there have been measurements of charge asymmetry fluctuations motivated by CME predictions from the STAR \cite{:2009uh} and PHENIX \cite{Ajitanand:2010rc} Collaborations at RHIC as well as from the ALICE \cite{Abelev:2012pa} at LHC. The precise meaning of these data is under investigations \cite{Bzdak:2009fc}.

In this contribution, I will discuss a number of recent progresses in the study of strong field effects in heavy ion collisions: 1) the Chiral Magnetic Wave (CMW) and its experimental manifestation via splitting of $\pi^{\pm}$ elliptic flow~\cite{Burnier:2011bf}; 2) the event-by-event azimuthal fluctuations of strong EM fields~\cite{Bloczynski:2012en} and its correlations with matter geometry; 3) a new mechanism for generating axial current in external electric field, the Chiral Electric Separation Effect (CESE)~\cite{Huang:2013iia}.

\section{The Chiral Magnetic Wave}

Since an external $\vec B$ field can induce both CME (\ref{cme}) and CSE (\ref{cse}), it is interesting to see the interplay between the two. As first pointed out by Kharzeev and Yee ~\cite{Kharzeev:2010gd},  the combination of the CME and CSE leads to a collective excitation in QGP called Chiral Magnetic Wave (CMW). Intuitively it is not difficult to understand how such a wave arises by analogy with the electromagnetic wave in which the varying $\vec E$ and $\vec B$ fields mutually induce each other. For the CMW in external $\vec B$ field, a fluctuation in the vector(axial) density $\mu_V$($\mu_A$) will induce axial(vector) three current $\vec j_A$($\vec j_V$) along $\vec B$ via CSE(CME) correspondingly, and such three currents will transport the charge densities (as per continuity equations) along $\vec B$ which induce further currents. At the end there will be two waves propagating along $\vec B$ for the vector and axial charge densities respectively. These two waves can also be linearly recombined into two waves for the left-handed and right-handed densities via $j_{R/L}=j_V\pm j_A$. Mathematically the waves can be described by the following wave equation:
\begin{eqnarray}
\left(\partial_0 \mp{v} \partial_1 -D_L \partial^2_1 -D_T\partial^2_T\right) j^0_{L,R}=0, \label{cmw}
\end{eqnarray}
with $v=\frac{N_c e B \alpha}{2\pi^2}$ the velocity of the wave and $\alpha$ the susceptibility connecting charge density and chemical potential. The last two terms include also the dissipative effects due to diffusion of charge densities with $D_L$ ($D_T$) the longitudinal (transverse) diffusion constant. 
 
\begin{figure}
	\begin{center}
		\includegraphics[scale=0.8]{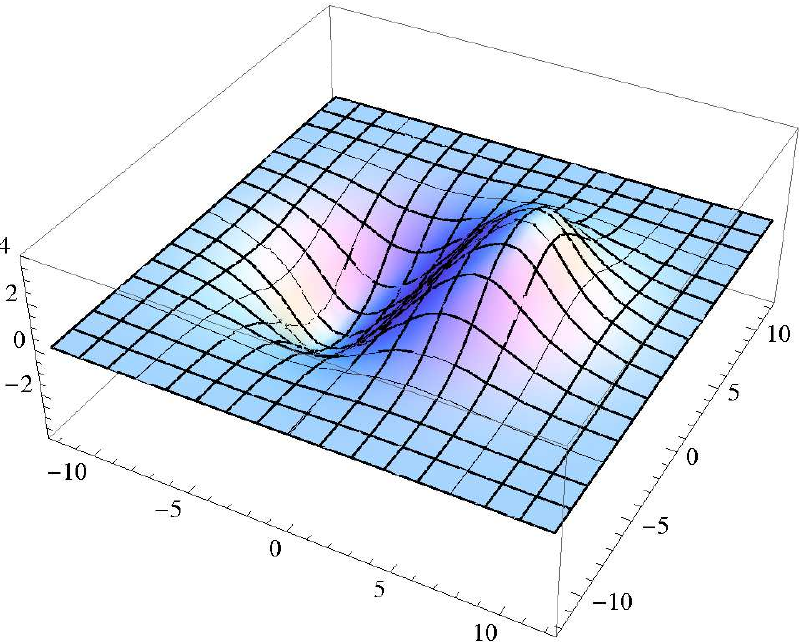}
			\includegraphics[scale=0.8]{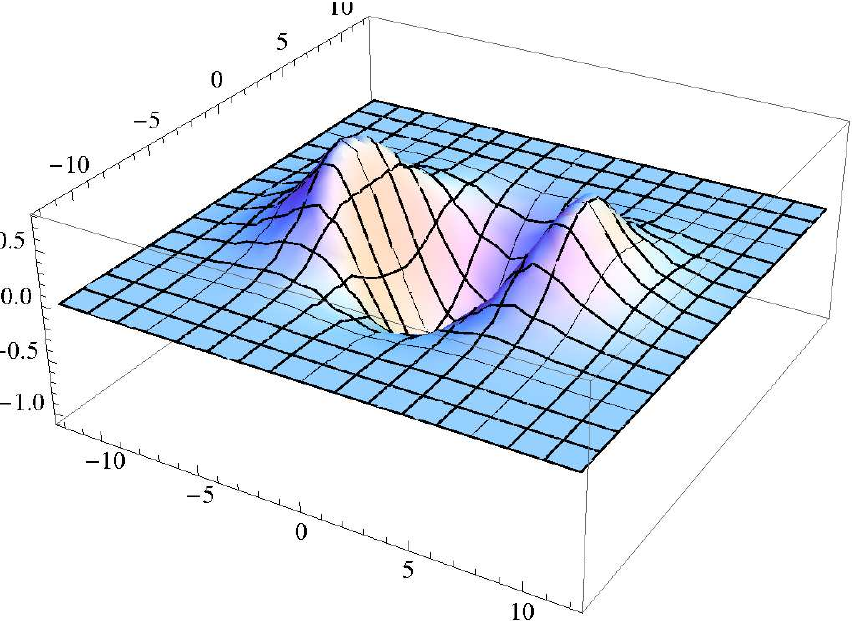}
		\caption{ Axial charge density (left) and electric charge density (right) in the plane transverse to the beam axis (computed with magnetic field strength  $eB=m_\pi^2$, lifetime of magnetic field $\tau=10$ fm,  temperature $T=165$ MeV, impact parameter $b=3$ fm).}
		\label{chiral_chem_pot}
		\vspace{-0.5cm}
	\end{center}
\end{figure}

Naturally one may wonder how the CMW may manifest itself in heavy ion collisions. It was first proposed in  \cite{Burnier:2011bf} that the CMW  induces an electric quadrupole moment of the created QGP. If the overlapping zone picks up nonzero vector charge density from the colliding nuclei, and starting with such density in external $\vec B$ field (along the out-of-plane direction), the CMW will transport both vector and axial charge densities, eventually leading to a dipole moment of axial density while a quadrupole moment of vector density (here the electric charge density), both aligned along the $\vec B$ direction. Note that such initial vector charge density on average becomes larger and larger with decreasing beam energy, while at high beam energy one can still have events with sizable initial vector density by fluctuations. By numerically solving the CMW equation (\ref{cmw}) with proper initial condition and  properties of QGP, it was shown in \cite{Burnier:2011bf} that the CMW leads to the axial charge dipole and electric charge quadrupole moments: see the plots in Fig.\ref{chiral_chem_pot}. 

\begin{figure}
\begin{center}
\includegraphics[scale=0.5]{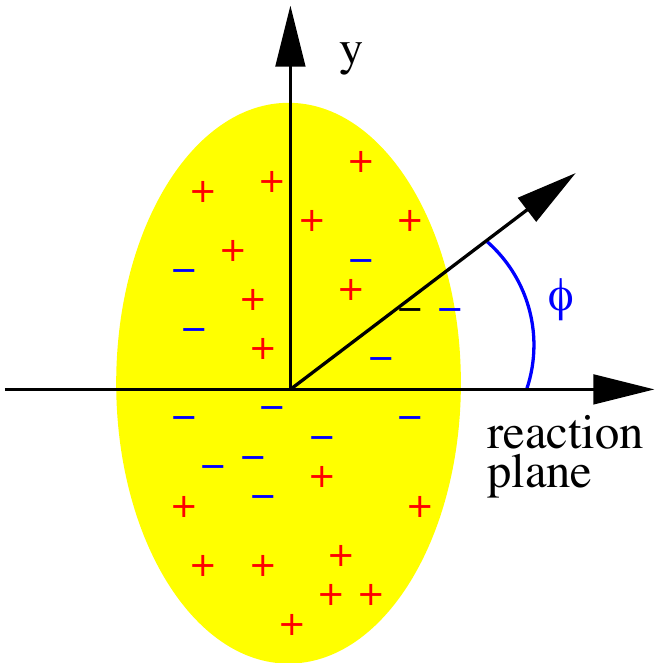} 
\includegraphics[scale=0.42]{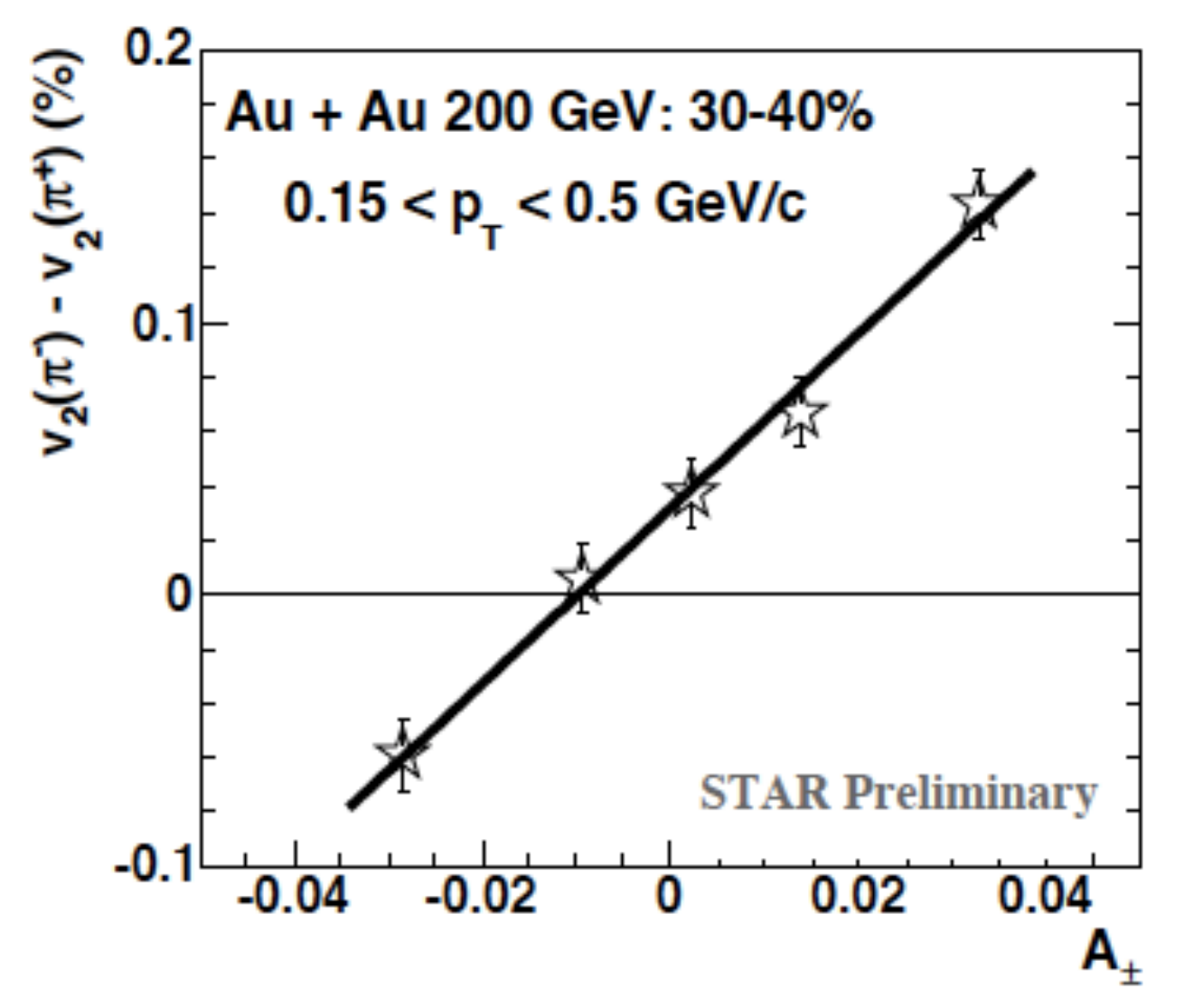} 
\includegraphics[scale=0.45]{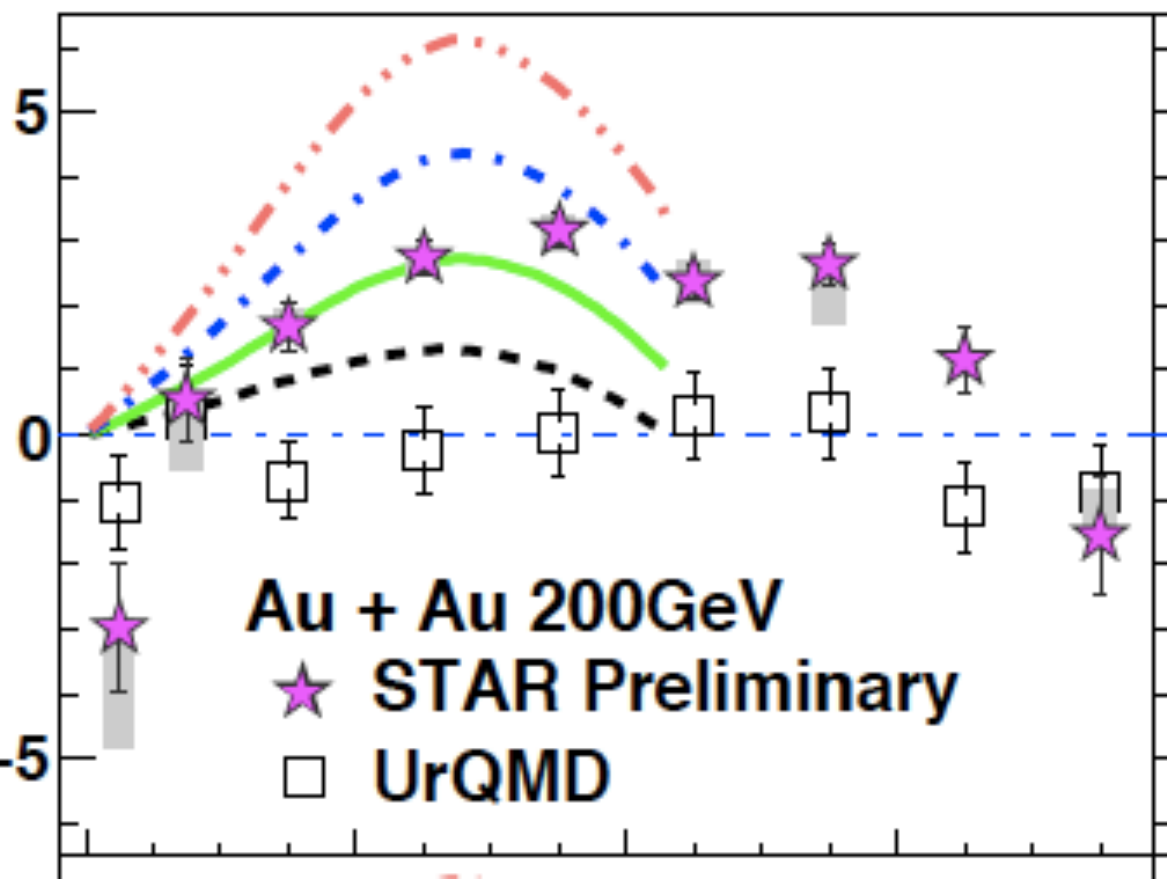}
\label{fig_v2}
\caption{ (left) Schematic demonstration of charge quadrupole being boosted by strong collective (radial) flow; (middle) elliptic flow splitting $v_2^{\pi^-}-v_2^{\pi^+}$ versus net charge asymmetry $A_{\pm}$ measured by STAR at $200\rm GeV$; (right) the slope parameter $r_e$(in \%) versus centrality, with black, green, blue, red lines from CMW calculations with magnetic field lifetime $\tau=3,4,5,6\, \rm fm/c$ respectively.}
\vspace{-0.5cm}
\end{center}
\end{figure}

It was further predicted in \cite{Burnier:2011bf} that such an electric charge quadrupole moment leads to a splitting between the positive/negative pions' elliptic flow. The idea is that such a spatial quadrupole charge distribution (at the end of the plasma phase) will be carried by strong radial flow and converted into azimuthal charge distribution in the final momentum space, resulting in more negative particles moving in-plane while more positive particles moving out-of-plane: see the demonstration in Fig.\ref{fig_v2} (left panel). This CMW-induced splitting between the $v_2$ of $\pi^\pm$ can be quantified by:
\begin{eqnarray}
v_2^- - v_2^+= r_e\,A \,\, . \label{v2split}
\end{eqnarray}
where the splitting is linear in the net charge asymmetry 
$A=\frac{\bar N_+-\bar N_-}{\bar N_+ +\bar N_-}$ with the slope $r_e$ being the quadrupole moment determined from net charge distribution due to the CMW evolution. Such a splitting was indeed first confirmed at low beam energies by STAR~\cite{Mohanty:2011nm}, in agreement with the prediction from \cite{Burnier:2011bf}. More recently the STAR has also systematically measured this $v_2$ difference as a function of net charge asymmetry at top RHIC energy~\cite{Wang:2012qs}, which shows a linear dependence on $A_{\pm}$ just as predicted in (\ref{v2split}): see the Fig.\ref{fig_v2} (middle panel). The magnitude of the extracted slope parameter $r_e$ and its centrality trend is also in good agreement with our CMW calculations~\cite{Burnier:2011bf}  assuming magnetic field lifetime $\tau=4 \, \rm fm/c $: see the Fig.\ref{fig_v2} (right panel). 

In short, we have shown that the CMW, stemming from interplay between CME and CSE, induces an electric charge quadrupole moment of QGP in heavy ion collisions and leads to a splitting of $\pi^{\pm}$ elliptic flow that linearly depends on the net charge asymmetry. Recent STAR measurements provide strong quantitative evidence of this effect from the CMW (noting though there are also proposals of other effects that may also contribute to  the $v_2$ splitting ~\cite{Steinheimer:2012bn}). To reach a final conclusion on the origin of the measured $v_2$ splitting, a number of improvements on the CMW-based modeling are underway, such as reducing uncertainty in the magnetic fields, better determination of the parameters in CMW model, more sophisticated simulation of the QGP evolution, etc. It will also be crucial to have more detailed measurements on the dependence of such $v_2$ splitting on beam energies, on particle $p_t$ and $\eta$, as well as on particle identities (e.g. pions versus kaons), which will be important for distinguishing different models.

\section{Azimuthal Fluctuations of Strong Fields in Heavy Ion Collisions}

\begin{figure*}
\begin{center}
\includegraphics[width=6.4cm]{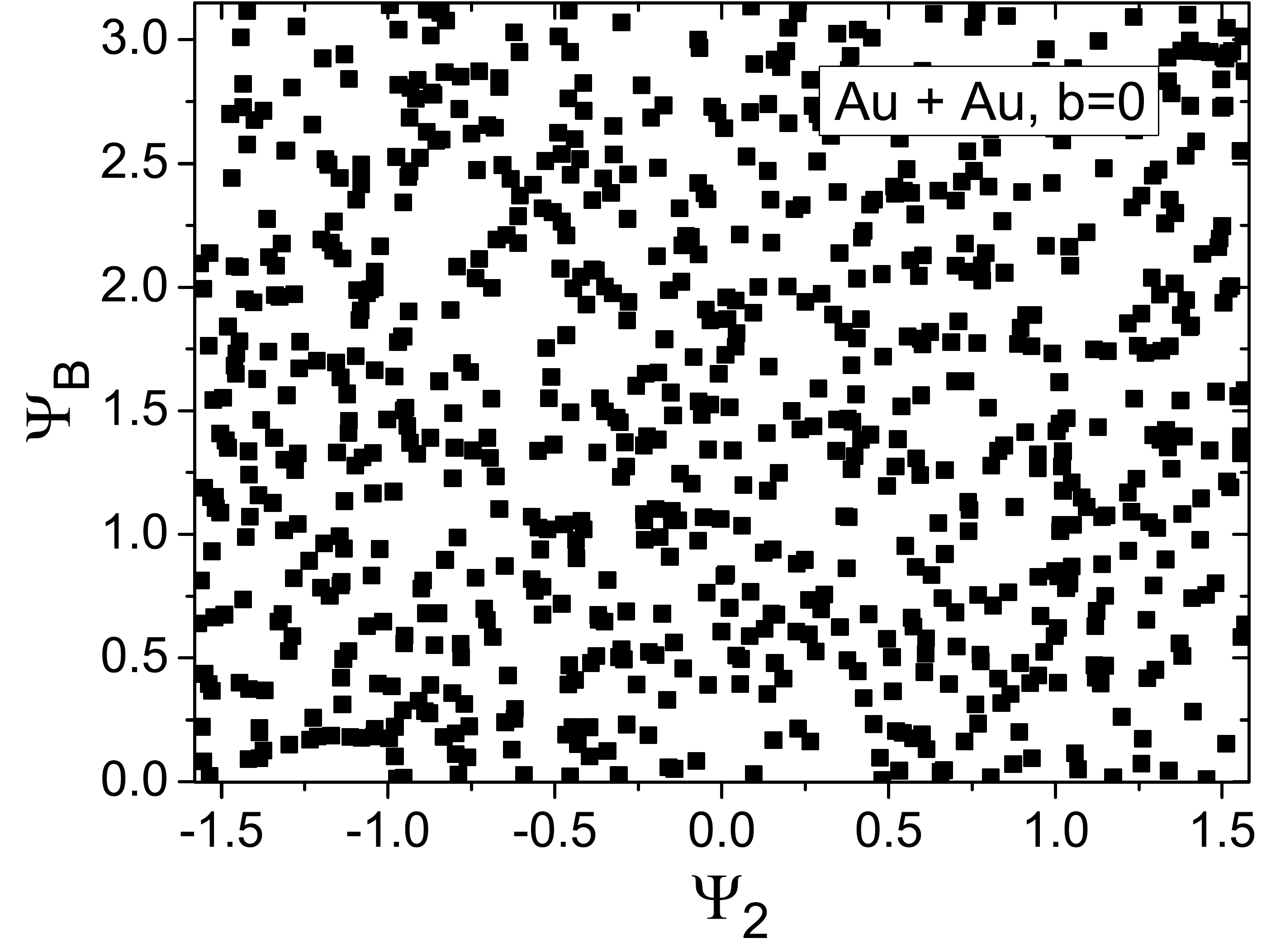}
\includegraphics[width=6.4cm]{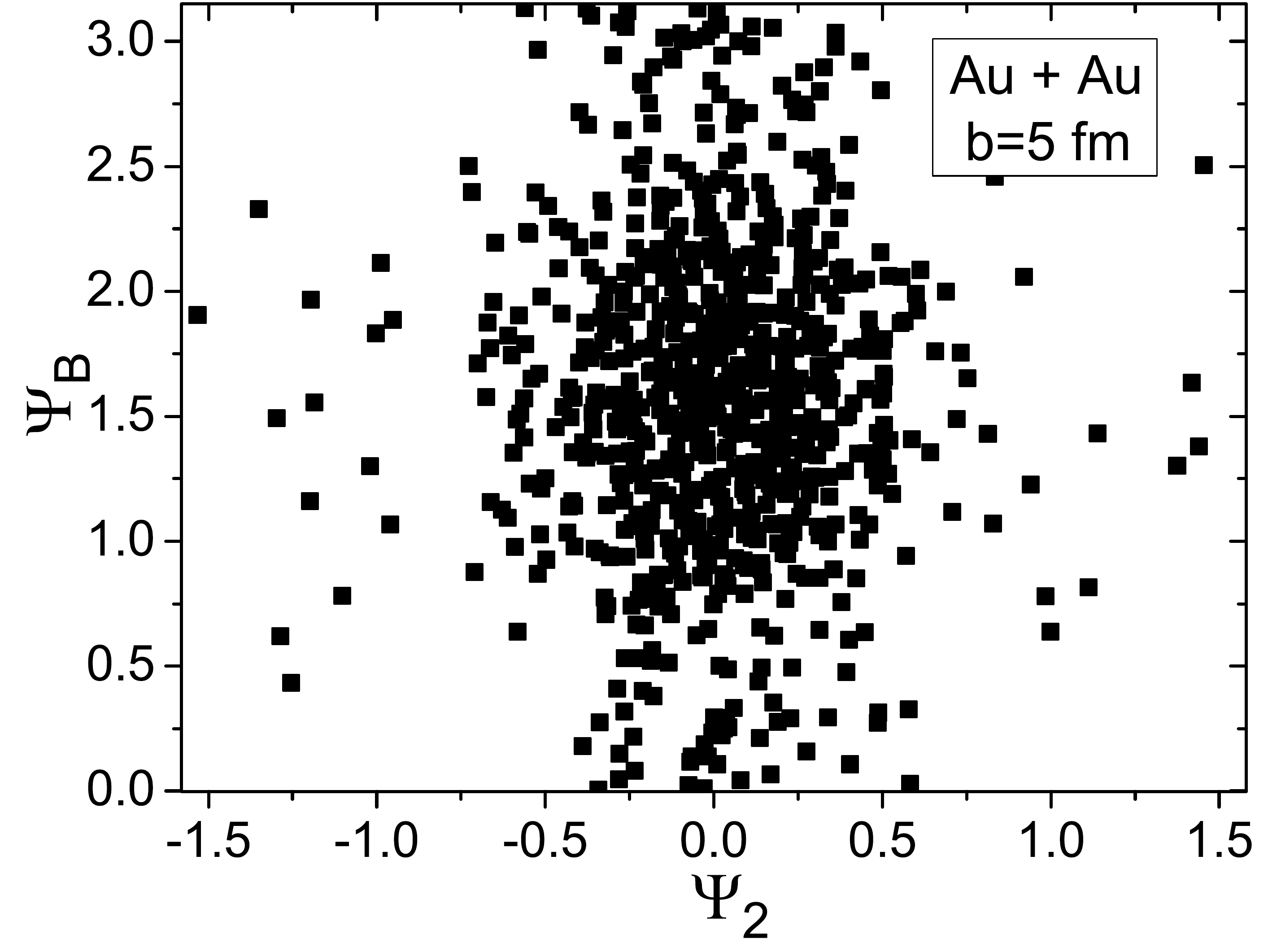}\\
\includegraphics[width=6.4cm]{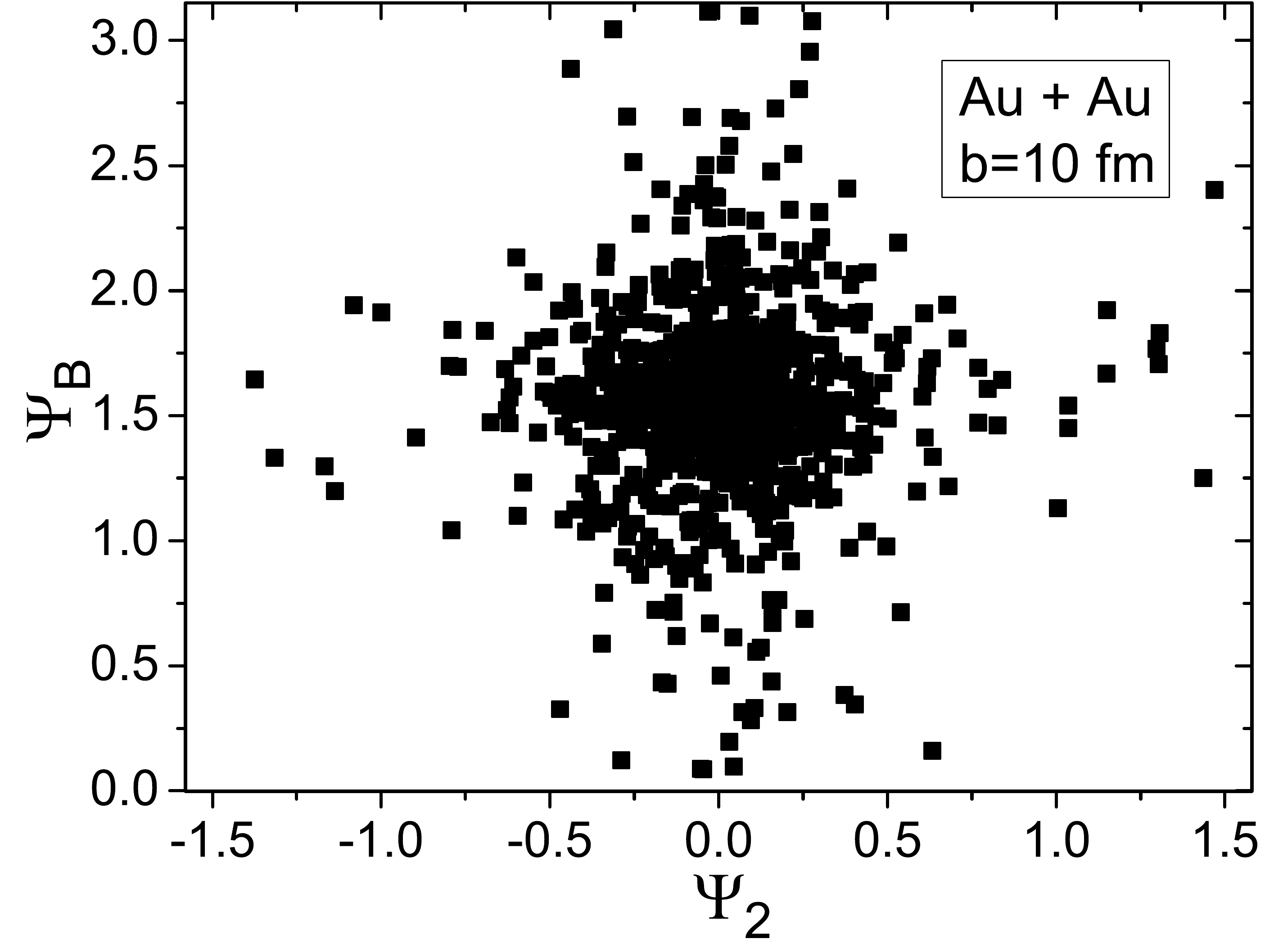}
\includegraphics[width=6.4cm]{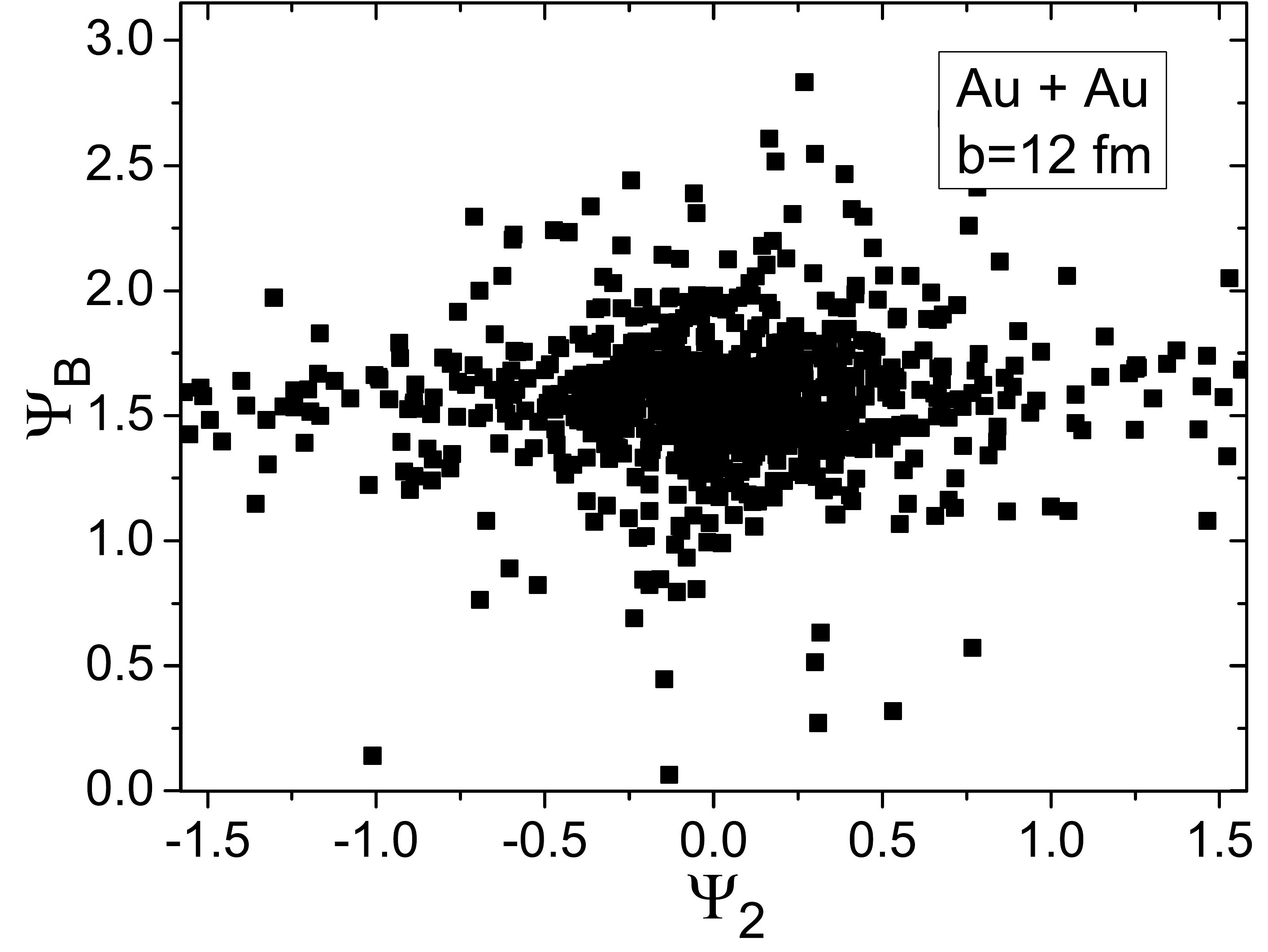}
\caption{The scatter plots on $\Psi_B$-$\Psi_2$ plane  at impact parameters $b=0, 5, 10, 12$ fm for Au + Au collision at RHIC energy. Here $\Psi_{\mathbf B}$ is the azimuthal direction of $\vec B$ field (at $t=0$ and ${\mathbf r}=(0,0,0)$) and
$\Psi_2$ is the second harmonic participant plane.}
\label{scat}
\end{center}
\vspace{-0.5cm}
\end{figure*}

The strong EM fields during the early stage of heavy ion collisions are the essential elements for the CME,CSE and CMW effects discussed here as well as for various other strong field effects studied in the literature ~\cite{other_effects}. In order to make comparison with experimental data, it is extremely important to fully quantify such EM fields. One big issue is its time evolution which critically depends on the medium  feedback to the fast decaying fields \cite{McLerran:2013hla,Tuchin:2013apa}. A conclusive answer may require more understanding of the pre-thermal partonic system~\cite{Blaizot:2011xf}, in particular the electric conductivity in such off-equilibrium environment. Another important factor is  the event-by-event fluctuations in the initial condition which are shown to bring sizable changes to the calculated magnitudes of these fields \cite{Bzdak:2011yy,Deng:2012pc}. What was not known before and was first studied by us in \cite{Bloczynski:2012en}, is the event-by-event azimuthal orientation of the EM fields with respect to the also fluctuating matter geometry in the same event. This information is absolutely essential for meaningful comparison between data and any effect induced by such fields.

\begin{figure*}
\begin{center}
\includegraphics[width=6.4cm]{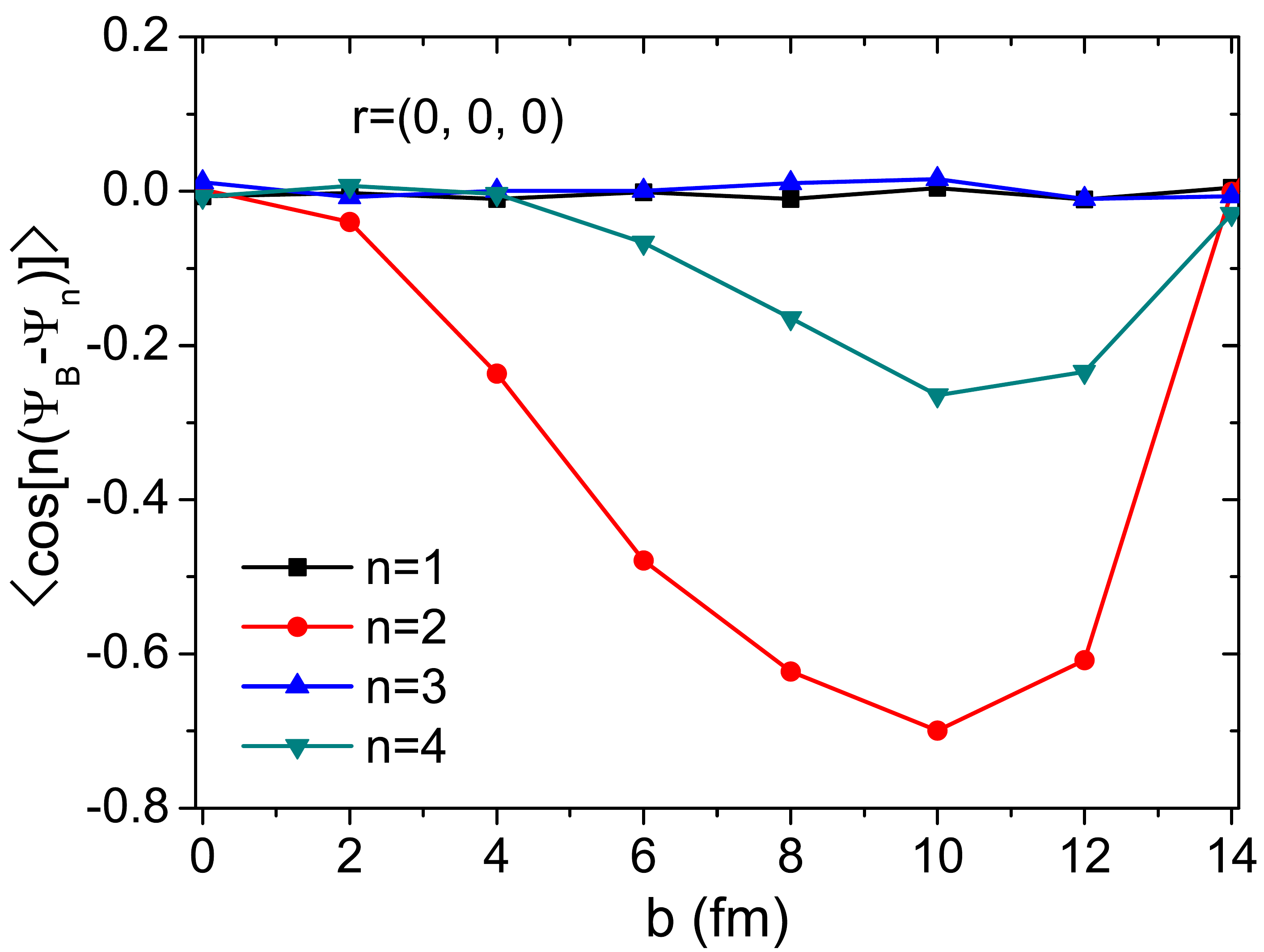}
\includegraphics[width=6.4cm]{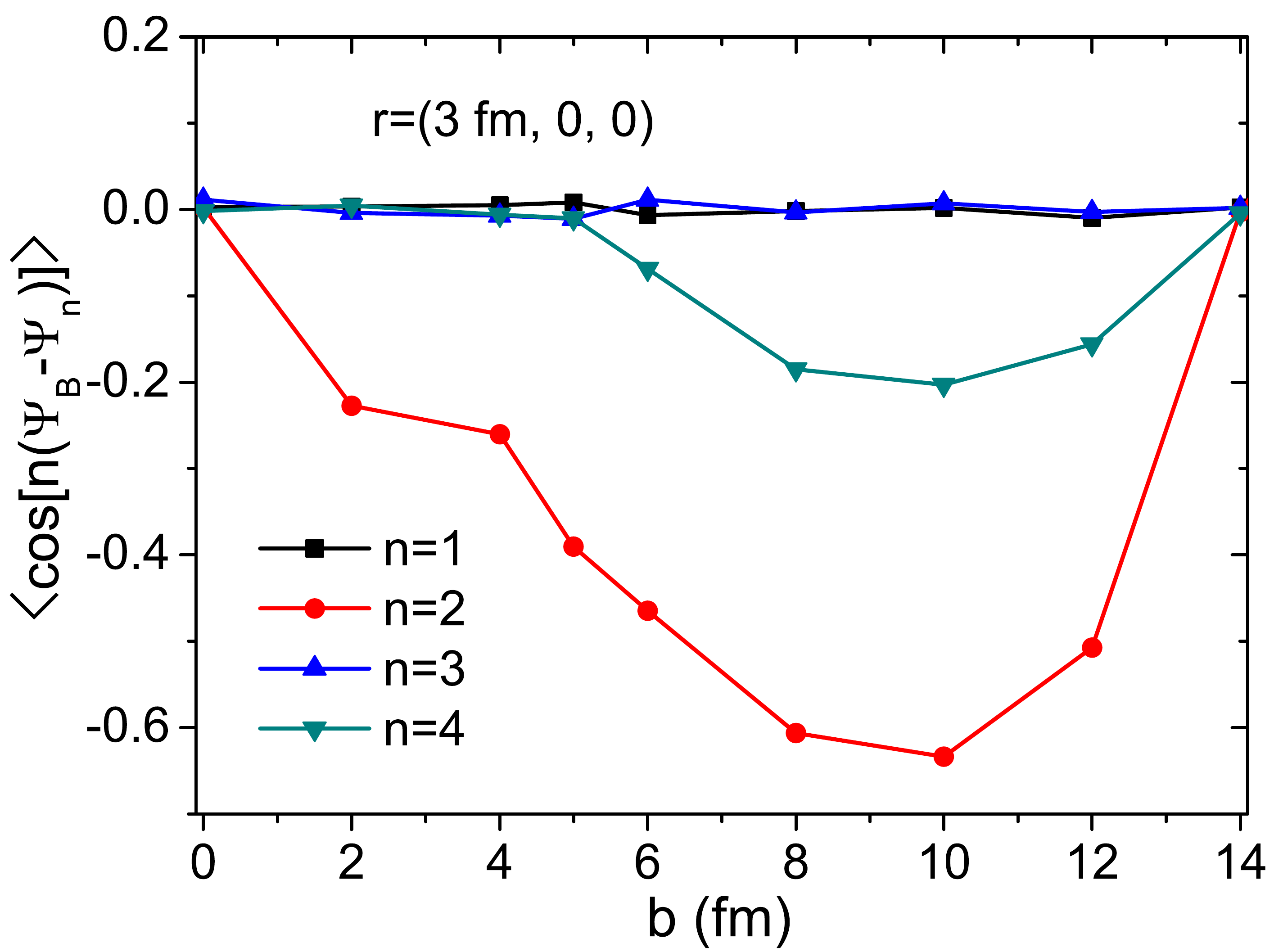}\\
\includegraphics[width=6.4cm]{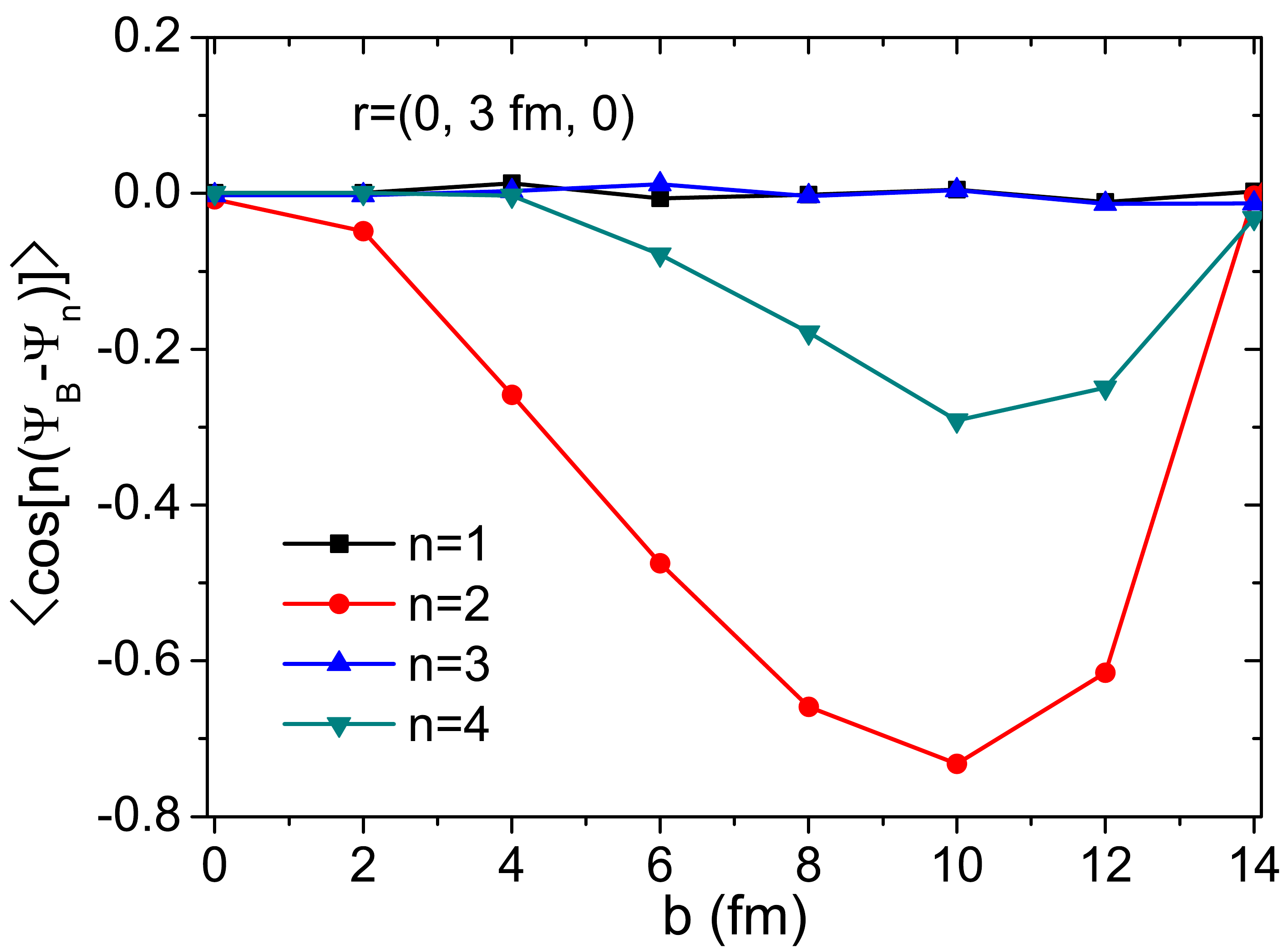}
\includegraphics[width=6.4cm]{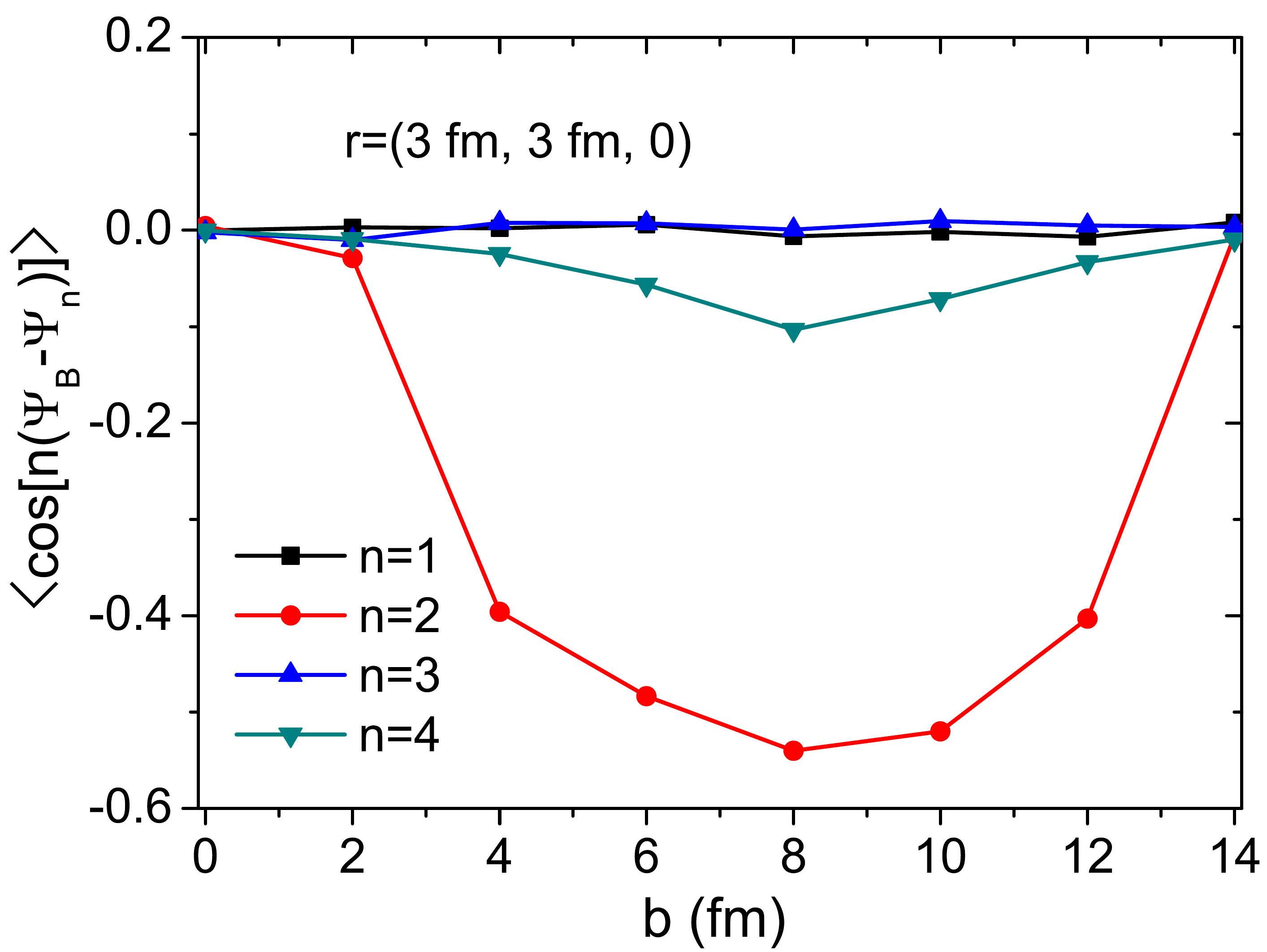}
\caption{The correlations $\langle\cos[n(\Psi_{\bf B}-\Psi_n)]\rangle$ as functions of impact parameter for $n=1,2,3,4$ at four different positions
on the transverse plane: (from left to right) ${\mathbf r}=(0,0,0)$ fm; ${\mathbf r}=(3,0,0)$ fm; ${\mathbf r}=(0,3,0)$ fm; ${\mathbf r}=(3,3,0)$ fm. }
\label{icos2}
\vspace{-0.5cm}
\end{center}
\end{figure*}

Let us consider a heavy ion collision event. Due to fluctuations the initial matter's azimuthal distribution can be lumpy and irregular, and can be decomposed into various harmonic ``participant planes'' characterized by certain angles $\Psi_n$ with respect to the ideal ``reaction plane''. Similarly in a given event the EM fields' transverse components $\vec B_\perp$ (and $\vec E_\perp$ as well) may also point toward some azimuthal direction $\Psi_{\bf B}$ other than that from the ``optical geometry''. What really matters for experimental measurements is the relative orientation of $\Psi_{\bf B}$ with respect to the matter geometry $\Psi_n$ in the very same event. Such correlations between $\Psi_{\bf B}$ and $\Psi_n$ fluctuate from event to event and typically bring in a reduction to the measured signals from the intrinsic strength of the signals (of various field induced effects) by the following factor:
\begin{eqnarray}
\label{reduct1}
R_n=\langle\cos(n\bar{\Psi}^n_{\bf B})\rangle=\langle\cos[n(\Psi_{\bf B}-\Psi_n)]\rangle.
\end{eqnarray}
It is therefore necessary to study such correlations  and to quantify the above factor.

In Fig.~\ref{scat} we show the scatter plots  from all events at given $b$ on
the $(\Psi_{\bf B}, \Psi_2)$ plane which visualize the 2D probability distribution density.  As one can see for the most central collisions $b=0$, the events are almost
uniformly distributed indicating negligible correlation between 
$\Psi_{\bf B}$ and $\Psi_2$. For $b=5, 10$, and $12$ fm, the event distributions evidently concentrate
around $(\Psi_{\bf B}, \Psi_2)=(\pi/2,0)$ indicating a correlation between the two. Going from $b=5$ to $10$ and to $12$ fm, the spread in $\Psi_{\bf B}$ keeps shrinking while the spread in $\Psi_2$ clear grows with larger $b$. This is because with increase $b$, the $\Psi_{\bf B}$ is mostly from the spectators whose number increases and bears less fluctuations while the $\Psi_2$ is determined by participants whose number decreases and fluctuates more. This implies a non-monotonic trend of the azimuthal correlations between the magnetic field and the matter geometry. Indeed this can be seen by examining the centrality dependence of the factor $R$ in (\ref{reduct1}). In Fig.~\ref{icos2}, we show the computed average values of  $\langle\cos[n(\Psi_{\bf B}-\Psi_n)]\rangle$ for varied centralities from event-by-event determination
of the $\bf B$-field direction $\Psi_{\bf B}$ (at several different spatial points) and the participants harmonics, $\Psi_n$, $n=1,2,3,4$. We find the $\Psi_{\bf B}$ are most strongly correlated with $\Psi_2$, visibly correlated with $\Psi_4$ while not correlated with $\Psi_{1,3}$. For the angular  correlation between $\Psi_{\bf B}$ and $\Psi_2$ (which in the optical limit would differ by $\pi/2$ with $R=-1$), it is smeared out significantly in the very central and very peripheral collisions while stays strong for middle-centrality collisions. This finding bears important 
 implications on observables related with ${\bf B}$-induced effects such as the charged pair azimuthal correlation due to the CME, the
elliptic flow difference between $\pi^+$ and $\pi^-$ caused by the CMW,
and the soft photon emission from the conformal anomaly,  as thoroughly discussed in \cite{Bloczynski:2012en}. Our study suggests that the optimal centrality class for search of these strong field effects is that corresponding to impact parameter range $b\sim8-10$ fm.

\section{The Chiral Electric Separation Effect}

In this last part we report our recent finding of a new strong field effect called the Chiral Electric Separation Effect (CESE): the generation of axial current in external electric field~\cite{Huang:2013iia}.  It is useful to first remind ourselves of the long known Ohm's law that describes the generation of a vector current in conducting matter as a response to external electric field: 
\begin{eqnarray}
\vec j_V= \sigma \vec E, \label{con}
\end{eqnarray}
where $\sigma$ is the electric conductivity of the matter with the convention that the electric current is $e \vec j_V$. Putting this effect (\ref{con}) together with the CME (\ref{cme}) and CSE (\ref{cse}), one realizes that there is yet  one more possibility that has not been previously discussed, namely the possible generation of an axial current in the electric field. We find this indeed can occur when the matter has both nonzero vector and axial charge density i.e. nonzero $\mu_V$ and  nonzero $\mu_A$:
\begin{eqnarray}
\vec j_A = \chi_{e} \mu_V \mu_A \vec E , \label{cese}
\end{eqnarray}
which can be called a {\it Chiral Electric Separation Effect} (CESE). With this new relation found, one can nicely combine all four effects into the following compact form:
\begin{eqnarray}
\left(\begin{array}{c} \vec j_V \\ \vec j_A\end{array}\right) =  \left(\begin{array}{cc}
\sigma  & \sigma_5 \mu_A \\   \chi_e \mu_V \mu_A &  \sigma_5 \mu_V
\end{array}\right) \left(\begin{array}{c}
\vec E \\ \vec B
\end{array}\right). \label{vaeb}
\end{eqnarray}

To intuitively understand how the CESE (\ref{cese}) arises, let us consider a conducting system with chiral fermions. When an electric field is applied, the positively/negatively charged fermions will move parallel/anti-parallel to the $\vec E$ direction and both contribute to the total vector current as in Eq.(\ref{con}). If $\mu_V>0$ then there will be more positive fermions (moving along $\vec E$) and if further $\mu_A>0$ then there will be more right-handed fermions than the left-handed ones: the net result will thus be a net flux of right-handed (positive) fermions moving parallel to $\vec E$. This picture is most transparent in the extreme situation when the system contains only right-handed fermions (i.e. in the limit of $\mu_V=\mu_A>0$), with both a vector and an axial current concurrently generated in parallel to $\vec E$. The same conclusion can be made when both $\mu_V$ and $\mu_A$ are negative.
  In cases with $\mu_V>0>\mu_A$ or $\mu_V<0<\mu_A$  there will be an axial current generated in anti-parallel to the $\vec E$ direction. An explicit example is given in \cite{Huang:2013iia} for thermal QED plasma, with the computed CESE conductivity to be 
\begin{eqnarray}
\label{sigma_e}
\sigma_e =\chi_e \mu_V \mu_A  \approx20.499\frac{\mu_V\mu_A}{T^2}\frac{T}{e^3\ln(1/e)} \,\, .
\end{eqnarray}

As seen in Eq.(\ref{vaeb}), with the presence of external electromagnetic fields, the vector and axial densities/currents can mutually induce each other and get entangled together. In particular the new CESE effect introduces nonlinearity (aka the $\mu_V\mu_A$ term) and makes the problem more nontrivial. An linearized analysis of the  coupled evolution of the small fluctuations in vector and axial charge densities in the static and homogeneous external $\vec E, \vec B$ fields shows that in addition to the CMW modes, there are new collective excitations such as the Chiral Electric Wave (CEW) as well as vector and axial density waves: see more details in  \cite{Huang:2013iia}. 

Finally we point out that the CESE may induce specific charge azimuthal distribution pattern for created QGP in Cu-Au collisions at RHIC where there is strong in-plane electric field pointing from Au to Cu \cite{Hirono:2012rt}. Such pattern could in principle be measured.



\end{document}